\documentclass[]{spie}  %>>> use for US letter paper
\usepackage{amsmath,amsfonts,amssymb}
\usepackage{graphicx}
\usepackage[colorlinks=true, allcolors=blue]{hyperref}

\title{Observational Artifacts of NuSTAR: Ghost-rays and Stray-light}

\author[a]{Kristin K. Madsen}
\author[b]{Finn E. Christensen}
\author[c]{William W. Craig}
\author[a]{Karl W. Forster}
\author[a]{Brian W. Grefenstette}
\author[a]{Fiona A. Harrison}
\author[a]{Hiromasa Miyasaka}
\author[a]{Vikram Rana}

\affil[a]{California Institute of Technology, 1200 E. California Blvd, Pasadena, USA}
\affil[b]{DTU Space, National Space Institute, Technical University of Denmark, Elektronvej 327, DK-2800 Lyngby, Denmark}
\affil[c]{Space Sciences Laboratory, University of California, Berkeley, CA 94720, USA}

\authorinfo{Further author information: (Send correspondence to K.K.M.)\\K.K.M.: E-mail: kkm@caltech.edu, Telephone: 1 626 395 6634}

\newcommand{\as}{$''$}
\newcommand{\am}{$'$}
\newcommand{\nustar}{\textit{NuSTAR}}
\newcommand{\xmm}{\textit{XMM-Newton}}
\newcommand{\chandra}{\textit{Chandra}}
\newcommand{\degree}{$^\circ$}

\begin{document} 
\maketitle

\maketitle

\begin{abstract}
The \textit{Nuclear Spectroscopic Telescope Array} (\nustar), launched in June 2012, flies two conical approximation Wolter-I mirrors at the end of a 10.15\,m mast. The optics are coated with multilayers of Pt/C and W/Si that operate from 3--80\,keV. Since the optical path is not shrouded, aperture stops are used to limit the field of view from background and sources outside the field of view. However, there is still a sliver of sky ($\sim$1.0--4.0\degree) where photons may bypass the optics altogether and fall directly on the detector array. We term these photons Stray-light. Additionally, there are also photons that do not undergo the focused double reflections in the optics and we term these Ghost Rays. We present detailed analysis and characterization of these two components and discuss how they impact observations. Finally, we discuss how they could have been prevented and should be in future observatories.
\end{abstract}

% Include a list of keywords after the abstract 
\keywords{NuSTAR, Optics, Satellite}

\section{Introduction}
\label{sec:intro}  % \label{} allows reference to this section
\begin{figure}[h]
\begin{center}
\includegraphics[width=0.95\textwidth]{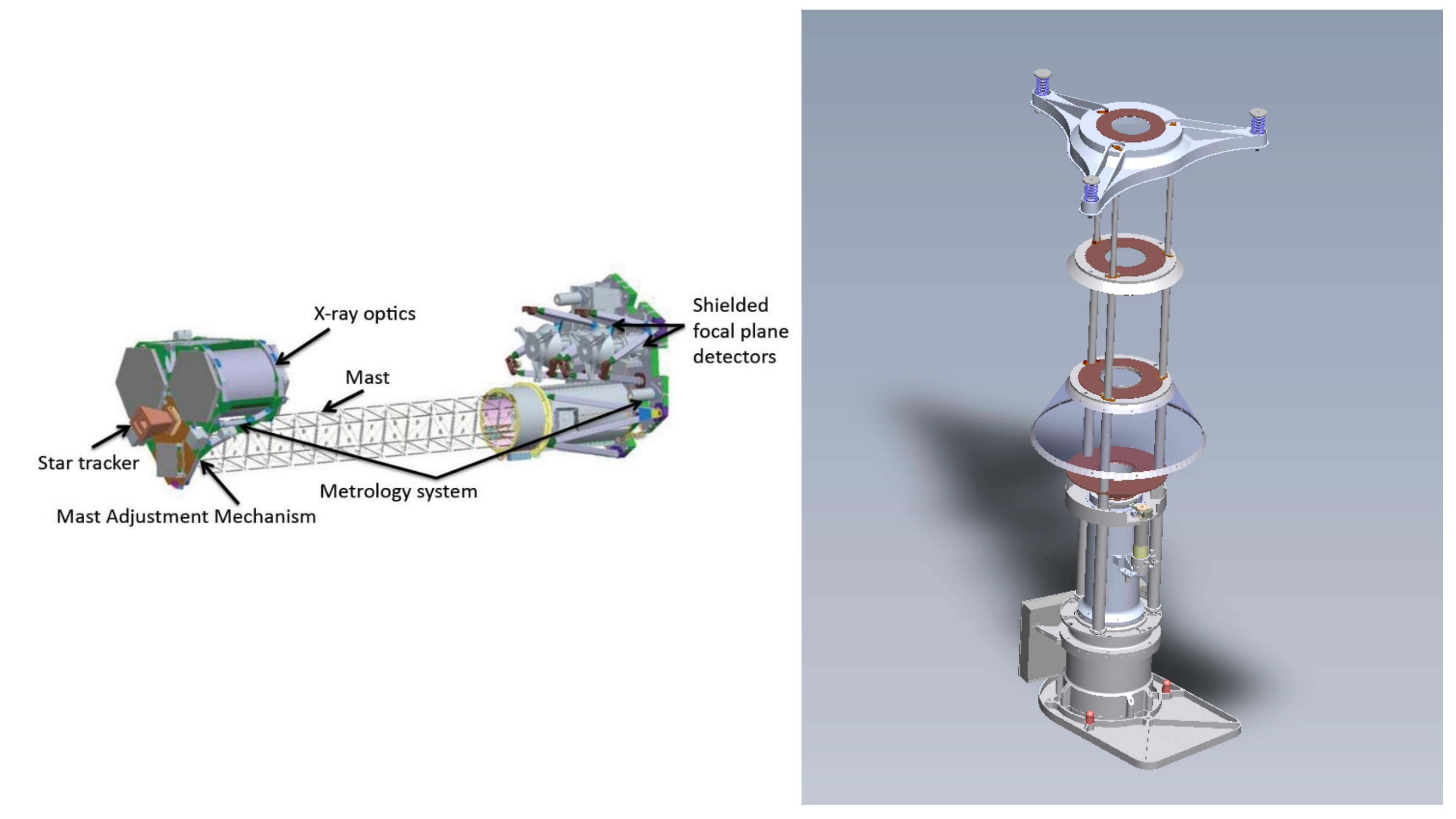}
\end{center}
\caption{ \textbf{Left}: Schematic of \nustar. The dimension of the mast is not to scale. \textbf{Right}: Schematic of the focal plane and the aperture stop assembly.}
\label{instrument}
\end{figure}

The \textit{Nuclear Spectroscopic Telescope Array} (\nustar), launched in June 2012 \cite{Harrison2013}, is a focusing X-ray observatory operating in the energy range 3--80\,keV. The schematic of \nustar\ is shown in Figure \ref{instrument} (\textit{left}). It carries two co-aligned conical Wolter-I approximation \cite{Petre1985} optic modules that focus onto two identical focal plane modules called FPMA and FPMB. The optical and focal plane benches are separated by a mast providing a focal length of 10.15\,m. A laser metrology system is used to keep track of the lateral and angular displacement of the two benches \cite{Harp2010}, which is caused by thermal motions in the mast, and a star tracker co-aligned with the optics provides the absolute aspect \cite{Forster2016}. \nustar\ is a pointed observatory and averages about one observation a day.

\begin{figure}[h]
\begin{center}
\includegraphics[width=0.95\textwidth]{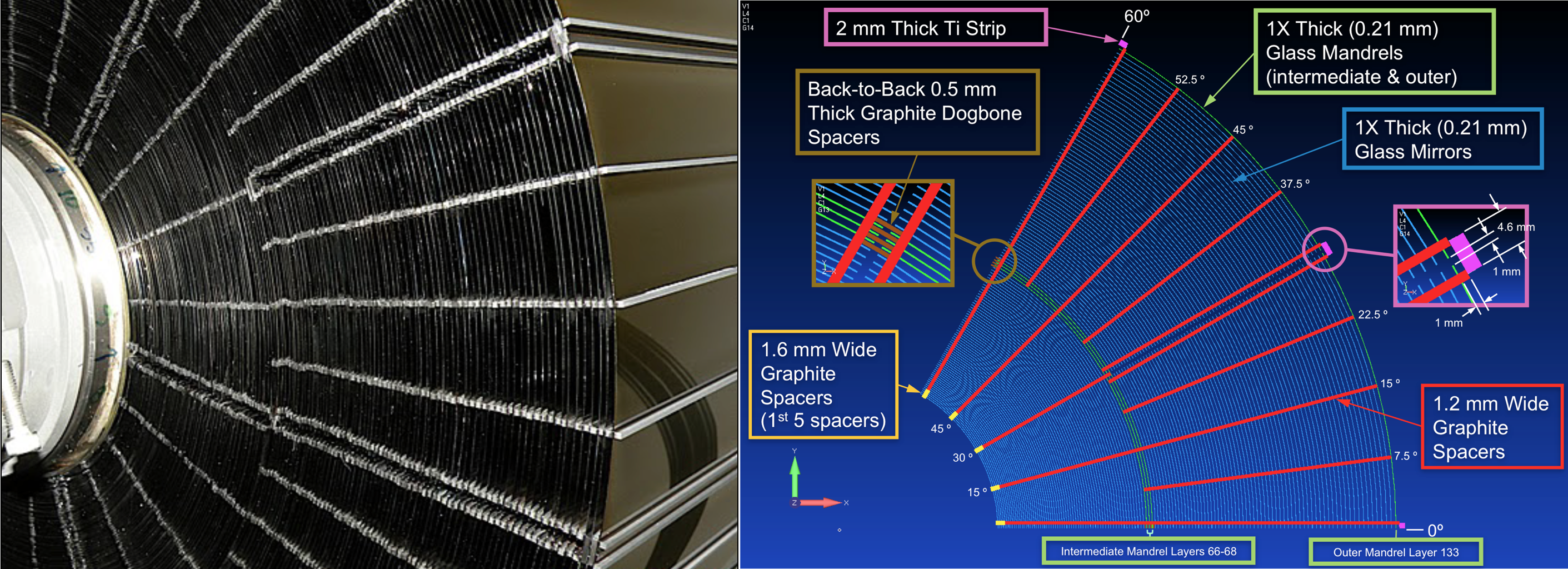}
\end{center}
\caption{ \textbf{Left}: Photo of a \nustar\ optic showing the structure of the mirror module and the graphite spacers holding the layers together. \textbf{Right}: Schematic of the a 60\degree\ wedge of the optics. The double `street' of spacers every 30\degree\ bounds a gap with no glass. This is the entry point of the 'streak' discussed in section \ref{section:streak}. The support spiders, not shown here, obscure every other one of these gaps. }
\label{optics}
\end{figure}

In a regular Wolter-I geometry \cite{Wolter1,Wolter2} the primary mirror is a paraboloid and the secondary a hyperboloid. In a Wolter-I approximation the two surfaces have been replaced with conical sections that reduces the complexity of the build, but comes at a cost of a larger Point Spread Function (PSF). Both constructions allow for very shallow grazing incidence angles that improve the efficiency of reflection for reasonable focal lengths. One single mirror, however, presents a very small area, and to achieve a greater area multiple mirrors are typically nested as shown in Figure \ref{optics} (\textit{right}). The spacing of the shells impacts the geometrical area; a looser filling factor utilizes more of the geometrical area, but the extra spacing makes it possible for photons that only undergo one reflection, or none at all, to make it through to the focal plane, while a denser filling reduces this amount of non-focused reflections, but also reduces the area due to self shadowing of the shells. The \nustar\ design favors a denser shell packing to reduce the non-focused reflections, but not all of them can be completely eliminated.

Over the course of the first four years of operation, we have acquired very good knowledge of these non-focused optical artifacts, which we collectively term Ghost Rays (GR), and present in section \ref{sec:ghostrays} the analysis of their character. We compare the observations to raytrace simulations and confirm that we understand the source of the artifacts.

Because of the penetrating hard X-rays, it is not feasible to shroud the optical path, and due to the open geometry \nustar\ is susceptible to stray-light (SL). The stray-light enters the detector aperture without having passed through the optics and undergone reflection. This component acts as an additional background \cite{Wik2014} and has implications for observatory planning. We present in section \ref{sec:straylight} detailed characterization of the component and its mitigation. 

In section \ref{sec:discussion} we summarize our findings and discus how these artifacts could have been avoided altogether, and how they should be avoided in future observatories.

\section{Detailed Instrument Overview}\label{section:overview}
The two \nustar\ optics were built to be geometrically identical. The length of each conical mirror section is 227\,mm, and the inner radius of the optics, where the two sections intersect, is 54.4\,mm and the outer 191.2\,mm. To achieve a high geometic area, \nustar\ has 133 nested shells; the outer 43 shells are coated with a W/Si multilayer, while the inner 90 shells are coated with Pt/C, limiting the highest efficient reflective X-ray energy to the Pt 78.4 keV K-edge. A multilayer is a stack of two alternating material pairs, called a bilayer, and in the case of \nustar\ the bilayer thicknesses are graded with thicker layers at the top for low energy reflection and thinner layers at the bottom for the high energy reflection \cite{Joensen1995,Madsen2009}. The angle of the innermost shell with respect to the optical axis is $\alpha_{primary,shell=1}=$1.342\,mrad (4.6\am) and  $\alpha_{p,133}=$4.715\,mrad (26.3\am) for the outermost shell. The angle of the secondary mirror is related to the primary by $\alpha_{s} = 3\alpha_{p}$. 

A section of the optics and its schematic is shown in Figure \ref{optics}. The optics are segmented and each shell is composed of multiple mirrors, mounted and held together by epoxied graphite spacers. The module can be divided into an inner and outer shell section that is separated by an intermediate mandrel, which is a strong-backed block of three shells (shells 66-68) \cite{Craig2011}. For the inner shells (1-65) there are 12 mirror segments and the spacers are positioned every 15\degree, while for the outer shells (69-133) there are 24 segments and the spacers are positioned every 12.5\degree. The span of the mirrors are also different between the inner and outer shells with 60\degree\ for the inner and 30\degree\ for the outer.

\begin{figure}[h]
\begin{center}
\includegraphics[width=0.95\textwidth]{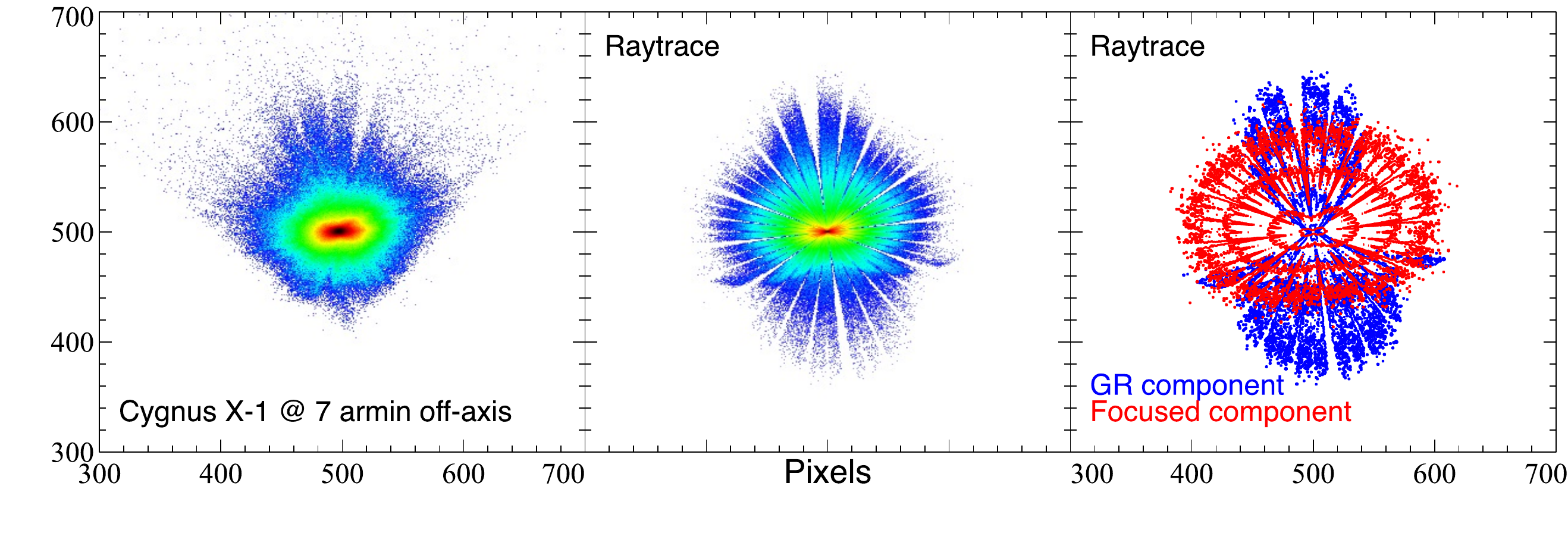}
\end{center}
\caption{\textbf{Left}: Off-axis observation of Cygnus X-1 (\nustar\ observation ID: 10014001001). The sharp edges at the bottom are the detector edges. \textbf{Middle}: Raytrace of the Cygnus X-1 observation. Detector extent not included. \textbf{Right}: Contours of the raytraced image are shown decomposed into the focused source component (properly double bounced photons) and the GR component to illustrate how different extraction regions centered on the source sample different amounts of the GR.}
\label{cygnusX1}
\end{figure}

The graphite spacers appear as dark absorption elements in the PSF (for example see Figure \ref{cygnusX1}). The gaps between mirror sections also appear as dark areas, because there is no mirror to reflect the photons. For specific angles, however, as discussed in section \ref{section:streak}, these light up with X-rays that arrive unobstructed at the focal plane. A support spider, which holds the optics in place within their cans, blocks out the mirror gaps every 60\degree\ and completely eliminates the mirror gaps of the inner shells.

Three aperture stops are attached to each focal plane module (FPM) as shown in Figure \ref{instrument} (\textit{right}). The top limiting aperture is located 833\,mm from the surface of the detectors and has a diameter of 58 mm. The total thickness of the solid part of the aperture is designed to be 1.88\,mm, layered into 0.75 mm Al, 0.13 mm Cu, and 1.0 mm Sn. These apertures act to limit the X-ray background and the stray-light from nearby sources. 

The field of view (FoV) is determined by the detector, which has dimensions 40$\times$40\,mm and results in a FoV of 13.5$\times$13.5\am. The physical pixel size is 604.8\,$\mu$m (12.3\as), and a factor of five sub pixel resolution is obtained in software for events sharing charge among multiple pixels to produce an effective pixel size of 2.5\as.

\section{Ghost Rays}
\label{sec:ghostrays}
\begin{figure}[h]
\begin{center}
\includegraphics[width=0.95\textwidth]{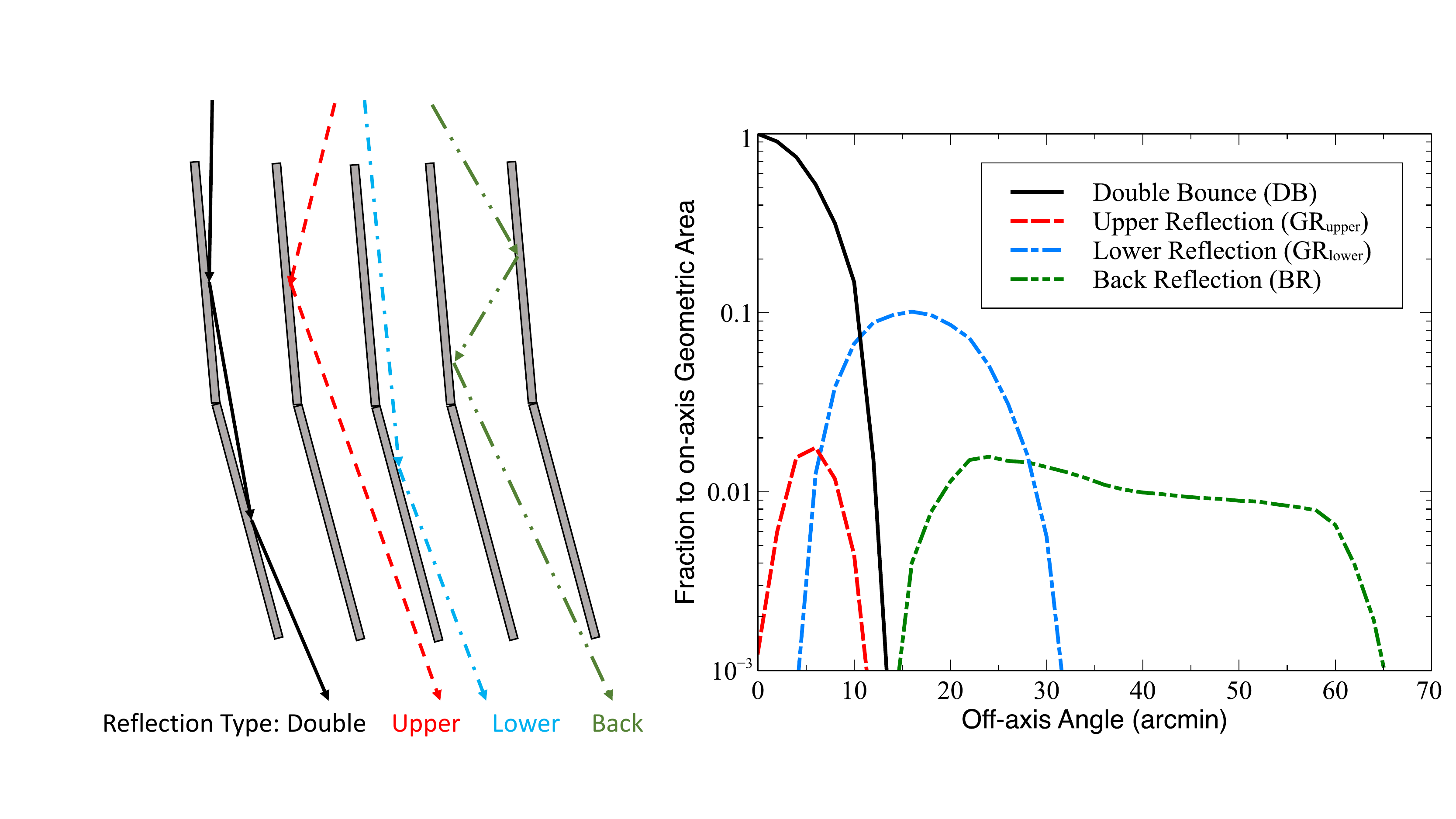}
\end{center}
\caption{\textbf{Left}: Illustration of different types of reflection in the optics. Black (solid) is the nominal double bounced reflection, which focuses the photons onto the focal plane. Red (dashed) and blue (dash-dot) are single reflection of the upper and lower conical mirror sections, collectively termed Ghost Rays (GR). The green (dash-dot-dot) ray illustrates the back reflection (BR) situation where a photon strikes the backside of the adjacent mirror, then reflects off the front side of the upper section and exists the optics. \textbf{Right}: The relative geometric area (not including reflectivity) of the various components in \nustar\ with respect to the on-axis geometric area. The areas have been corrected for aperture stop and the finite detector size.}
\label{GRillustration}
\end{figure}

Ghost Rays (GR) is the term given to the photons that only undergo a single reflection inside the optics. They either reflect from the upper (primary) or the lower (secondary) mirror section as shown in Figure \ref{GRillustration} (\textit{left}). In addition there is a back reflected component (BR), in which the photons strike the backside of the adjacent mirror first before reflecting off the upper mirror. 

\begin{figure}[h]
\begin{center}
\includegraphics[width=0.95\textwidth]{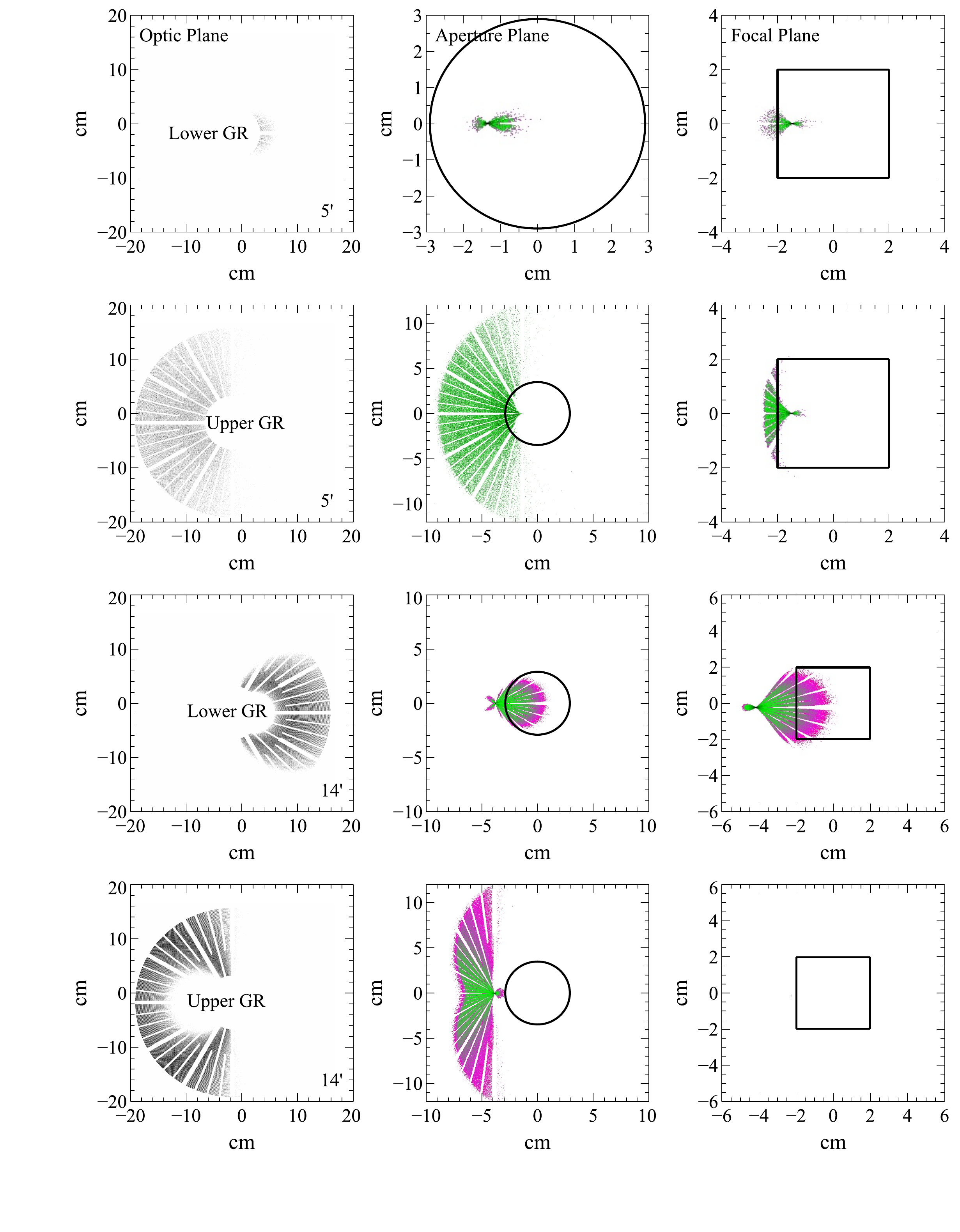}
\end{center}
\caption{The photon path of the upper and lower GR component from a source at 5\am\ and 14\am\ shown right after the photons have exited the optics (optic plane), at the location of the limiting aperture (aperture plane), and finally at the focal plane, for which we have removed all aperture stop rejected photons. The circle shows the extent of the aperture stop opening and any photons outside will be rejected. The square shows the extent of the focal plane detectors. Physically the source is moving along the positive x-axis, while its image is moving along the negative x-axis.}
\label{aperture_explained}
\end{figure}

The geometric area of all components, excluding the effects of reflection but including the aperture stop and the finite detector size, is shown in Figure \ref{GRillustration} (\textit{right}). The upper GR is the first component, and it is generated by photons that strike the upper mirrors at angles steeper than the nominal focusing graze angle. It dies out when the angle becomes so steep that the adjacent shell shadows it. The lower GR are made by photons arriving at angles that are shallower than the nominal graze angle. The lower GR reflections are thus produced on the opposite side of the mirror module than the upper GR as shown in Figure \ref{aperture_explained}. The Figure also demonstrates that the aperture stop is responsible for limiting the upper GR component. The lower GR component dies out when the photon angle becomes equivalent to the angle of the mirror. At this point the reflection of the backside takes over. Because the shells have different angles the components overlap. The true effective area, which includes the reflection, alters the areas for all components drastically as a function of energy. 

\begin{figure}[h]
\begin{center}
\includegraphics[width=0.95\textwidth]{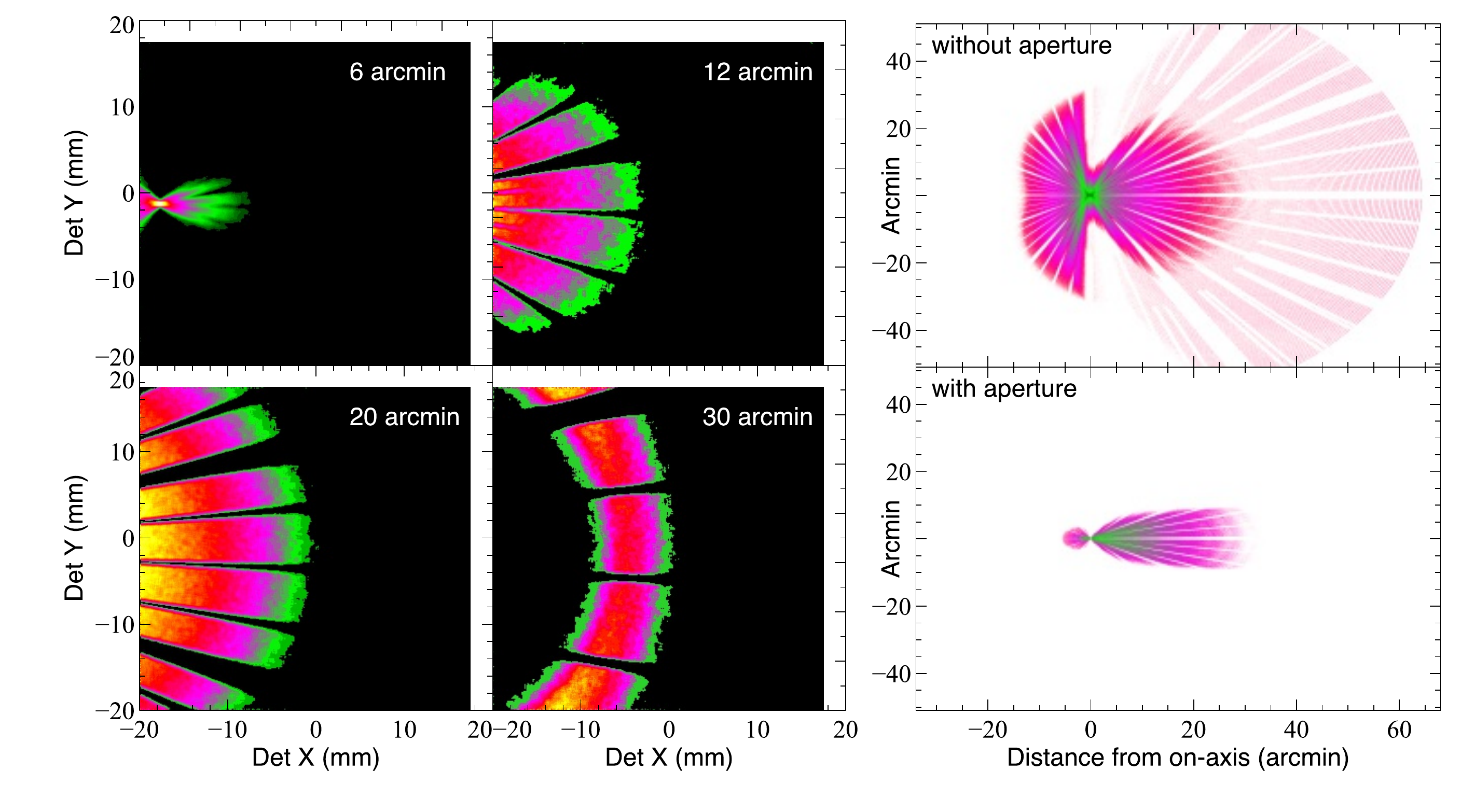}
\end{center}
\caption{\textbf{Left}: Ray-trace simulation of GR falling on the detectors at different off-axis angles. \textbf{Right}: Composite image of the GR pattern from a source for every 2\am\ for an infinite focal plane. Photons that slip straight through the optics without reflecting on either surface can be seen extending like a fan from 20--60\am. These and a large part of the GR are rejected by the aperture stop.}
\label{grimage}
\end{figure}

Figure \ref{grimage} shows raytraces of the GR component (\textit{left}) at several different off-axis angles as they would appear on the detector, and a composite image of the GR (\textit{right}) with and without the rejection of the aperture stop. This composite image does not include the BR component, and the fan feature extending from 20--60\am\ in the upper image are photons that would, if not for the aperture stop, have made it straight through the optic without reflecting off any surface and have reached the detector.

The GR component is axisymmetric along the line from the source to the optical axis. The primary concern is that photons from a distant source will interfere with the analysis of a focused target source, but as shown there are often free regions on the detector, and ultimately it is the relative strengths and relative location of the two sources that decide whether the GR becomes a significant issue. 

\begin{figure}[h]
\begin{center}
\includegraphics[width=0.95\textwidth]{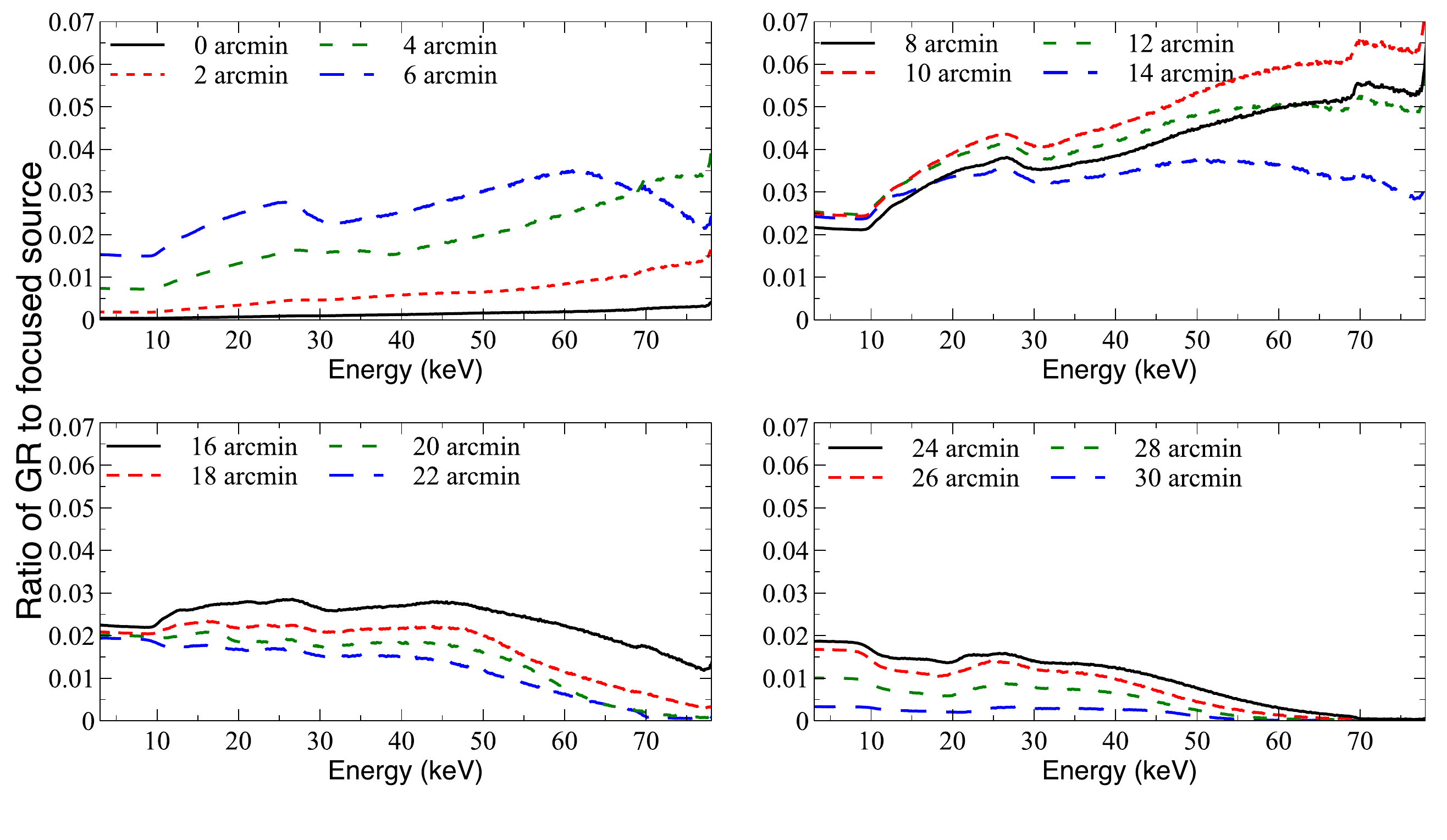}
\end{center}
\caption{Ratio of the total number of GR photons falling on the detector to the total number of source photons collected at the focal plane from its on-axis position. \textbf{In this raytrace the reflectivity component has been included}. As shown in Figure \ref{grimage} the GR illumination is not uniform across the detector and these curves are therefore not representative of an area average, but of the total spectrum collected from the entire focal plane. The curves show that small off-axis angles are dominated by the inner shells, which are predominantly responsible for the high-energy throughput, while with increasing off-axis angles the low-energy, outer shells, dominate.}
\label{grspectrum}
\end{figure}

Because of the non-uniformity, it is not practical to calculate the GR flux with respect to a typical extraction region. Instead we show in Figure \ref{grspectrum} the spectrum of the GR, including the reflectivity component, as collected from the entire detector as a fraction of the source spectrum had it been imaged on-axis. Below 10\,keV, the effective area remains close to the geometrical area due to the fact that most shells have grazing incidence angles less than the critical angle. The critical angle is the angle below which X-rays  undergo total external reflection. For angles less than the critical, the reflectivity is therefore close to 100\% and the majority of photons successfully reflected. This is seen as a flat spectrum, with only a very slight energy dependency, because the critical angle is changing across the shells. Above 10\,keV, reflection alters the spectrum of the GR. The inner shells, responsible for most of the high energy area, are rejected by the aperture stop with increasing off-axis angles (see Figure \ref{aperture_explained}) - and conversely the outer shells responsible for most of the mid range energy throughput are rejected by the aperture stop at smaller anglesc - which is why in Figure \ref{grspectrum} there is first an increase in the high energy part of the spectrum, then a decrease at higher off-axis angles. 

\subsection{Effective Area Corrections}
\label{sec:areacorr}
As demonstrated above, the GR component appears as early as 2\am\ off-axis, but only becomes significant above 3\am. In most cases, planned science targets are located within 2\am\ of the optical axis, but that is not always the case. In these instances a source may contaminate itself with its own GR. This leads to an increase in the source spectrum and because of the non-uniform nature of the GR it is not practical to treat it as an additional background. Instead we have modified the effective area to properly account for the GR inclusions.

While the GR increases the area, the aperture stop, responsible for reducing the diffuse background by limiting the FoV, also decreases the area with increasing off-axis angles due to the clipping of the edges of the optical path between detector and optic. This is primarily a geometrical effect, but because each reflecting shell has a different energy response, the selective rejection of different shells introduces a spectral dependency for both the aperture and the GR component, which is also subjected to the aperture correction. 

\begin{figure}[h]
\begin{center}
\includegraphics[width=0.95\textwidth]{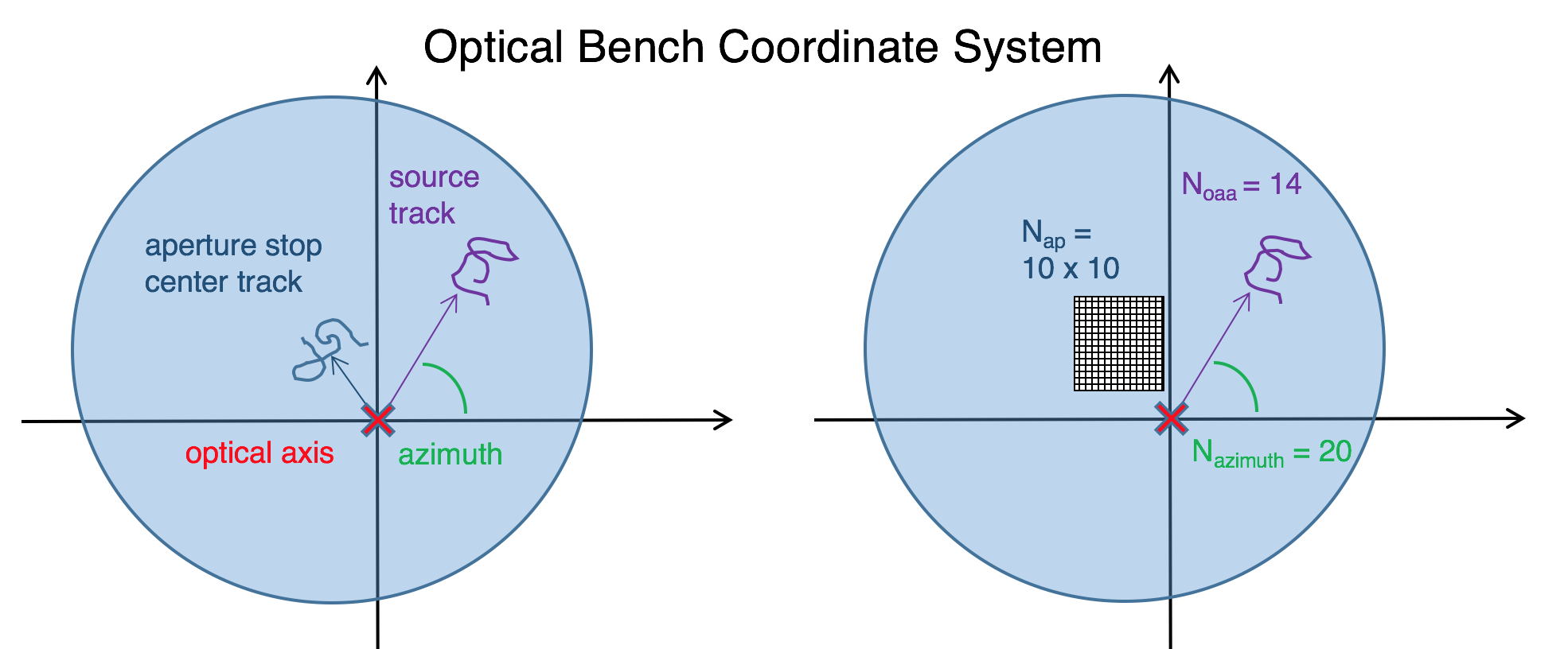}
\end{center}
\caption{GR and aperture stop correction schematic. The optics and their optical axis are stationary in the optical bench coordinate system. The circle represents the aperture stop and as a function of time we keep track of the center of the aperture with respect to the optics. The source moves along a different path caused by the spacecraft jitter. To cover all motions we raytrace 20 azimuth angles for every 18\degree, 14 off-axis angles binned every arcminute, and 10$\times$10 aperture stop positions at 1\,mm resolution.}
\label{grschematic}
\end{figure}

To complicate matters further, thermal gradients along the mast cause it to flex on an orbital time scale \cite{Forster2016}, and the motion smears out the PSF on the detector plane, resulting in a time dependent clipping of the area. However, the problem is completely determined by knowing the relative location of the aperture stop, stationary with respect to the detectors and the focal plane bench, with respect to the optical axis (OA) location, stationary in the optical bench frame. The two benches relative motions are tracked and reconstructed by a laser metrology system \cite{Harp2010}. Figure \ref{grschematic} lays out the geometry and time dependent terms. The coordinate frame is set in the optical bench frame, since this is where the OA is stationary. The circle represents the aperture stop and the blue track the motion of its center to the frame. The source, purple track, will be following its own motion based on the combined spacecraft jitter and mast motion. Because the aperture stop is not centered on the source, it is necessary to take azimuth into account as well. 

To generate the corrections we ran raytraces that covered a phase space of 20 azimuth angles for every 18\degree, 14 off-axis angles binned every arcminute, and 10$\times$10 aperture stop positions at 1\,mm resolution. We assumed a fixed PSF size for all energies with scattering parameters inside the raytrace adjusted to reproduce the polychromatic Hercules X-1 PSF. We did thus not take into account the secondary term of the a PSF varying in size as a function of energy \cite{An2014}. 

The above terms all go into calculating the aperture stop correction, but in addition to these the GR correction also requires the specification of an extraction region. This necessity is illustrated in Figure \ref{cygnusX1} where we have shown an off-axis observation of the bright accreting Black Hole X-ray binary Cygnus X-1 (\textit{left}), the raytraced observation (\textit{middle}), and the same raytrace where we have separated the focused photons from the GR component (\textit{right}). The amount the GR component contributes is dependent on the size of the extraction region, and due to the obvious complexity of the GR pattern the corrections were only derived for circular regions, which we include in steps of 20\as\ in radius. 

\begin{figure}[h]
\begin{center}
\includegraphics[width=0.95\textwidth]{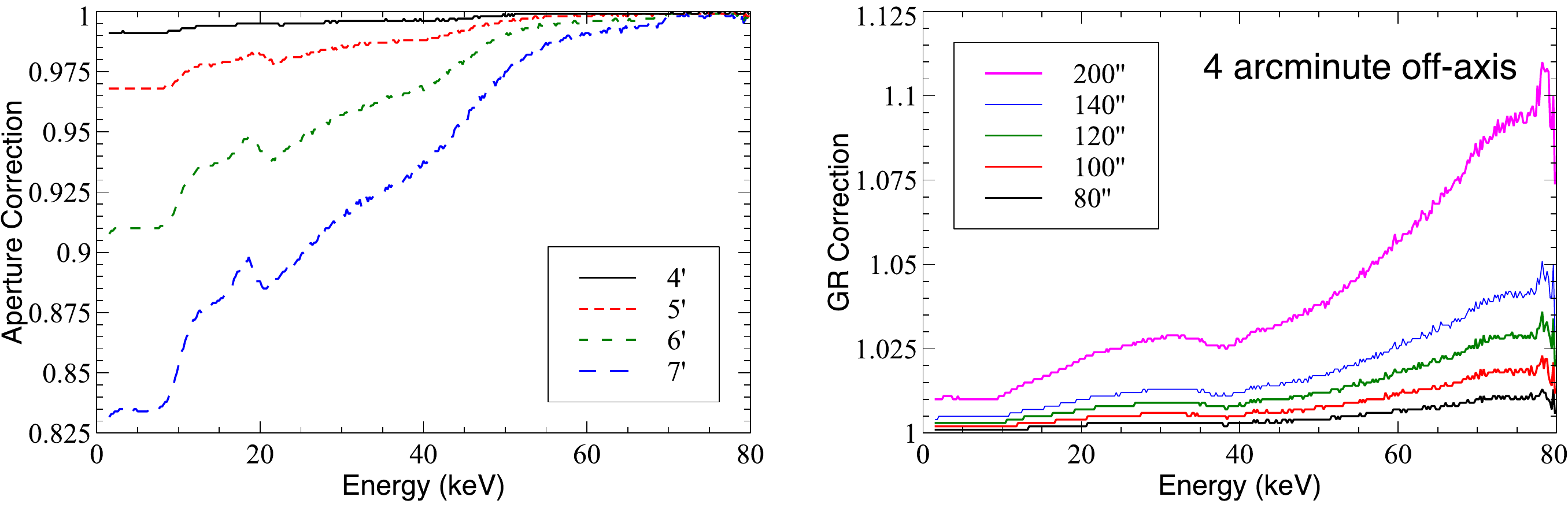}
\end{center}
\caption{\textbf{Left}: Correction to the effective area of proper double bounced photons due to the aperture stop. \textbf{Right}: Correction to the effective area due to the inclusion of GR photons for different extraction region sizes. The GR-correction also contains aperture corrections of the GR photons.}
\label{gr_ap_correction}
\end{figure}

An example of the magnitude of the aperture and GR corrections is shown in Figure \ref{gr_ap_correction}. The aperture correction is less than unity because it is removing photons from the effective area, and it is largest for low energy photons since the majority of these come from the outer shells of the optic and thus are more prone to being blocked by the aperture than light focused by the inner shells. The correction is only important for off-axis angles $>$ 3\am. For the GR corrections the net effect is an increased effective area.  The suppression of the low energies is because of the aperture stop correction to the GR. Both corrections are of a linear nature as a function of off-axis angle, and in practice we interpolate the correction tables when generating the observation specific instrument response.

\subsection{The `Streak'}\label{section:streak}
The `streak' is an artifact that is rare because it requires a fairly exact alignment of the source to the optics and detectors. It is caused by the absence of glass between mirror segments as shown in Figure \ref{optics} (\textit{right}), which allows the photons to pass right through the optics without reflecting off any surface and propagate down to the focal plane. This gap occurs at 60\degree\ intervals, and in order to reach the focal plane the source must be aligned azimuthally with one of the gaps and be within the correct off-axis angle. Since there are no gaps between shells with radius smaller than the intermediate mandrel, the smallest angle at which a photon can make it through to the focal plane is given by the radius of the intermediate mandrel at an R$\sim$108.7\,mm (shell 69), $\theta_{min} \sim 37$\am. The largest angle is determined by the size of the optics and is $\theta_{max} \sim 65$\am.

\begin{figure}[h]
\begin{center}
\includegraphics[width=0.95\textwidth]{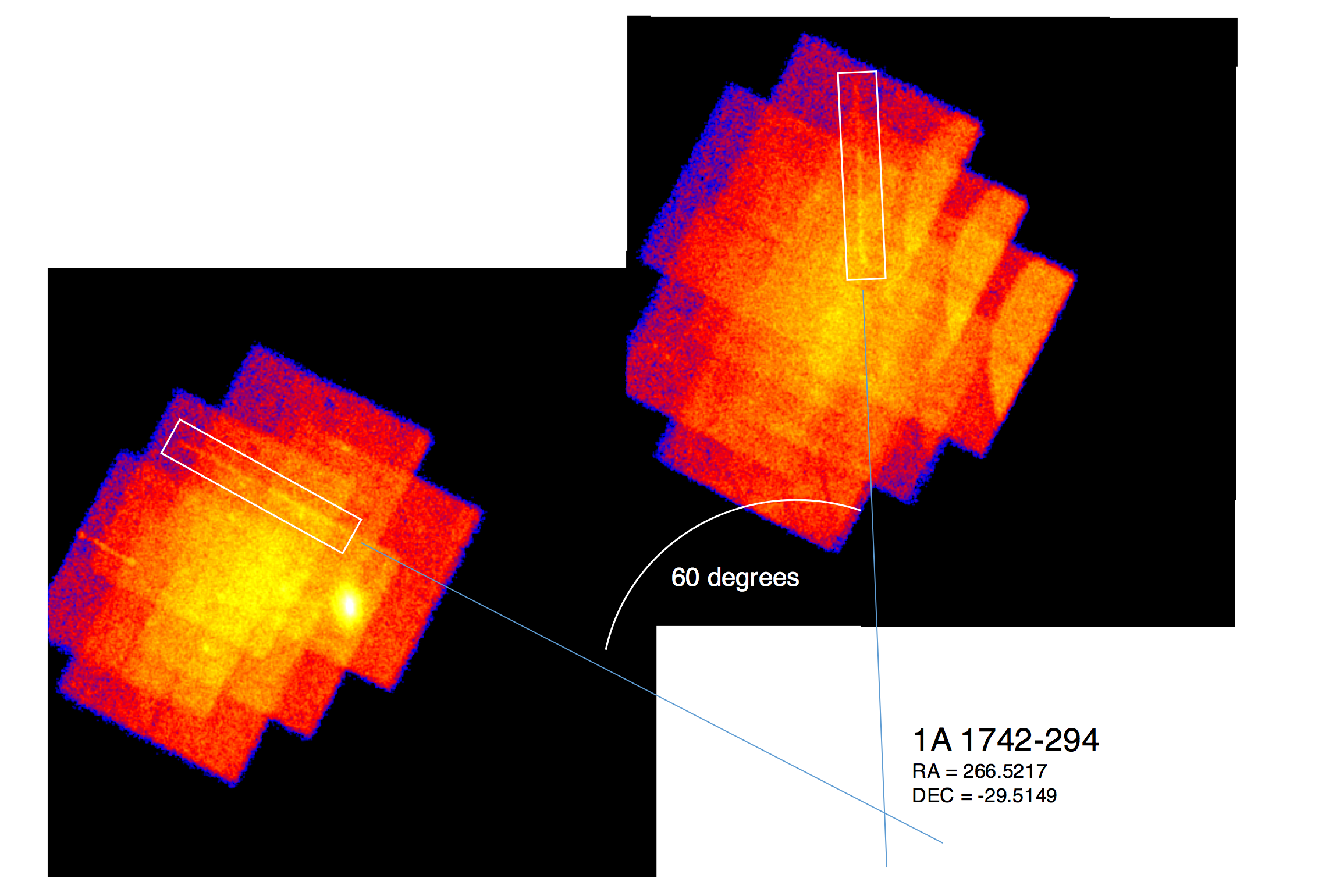}
\end{center}
\caption{During a Galactic Plane survey \cite{Hong2016} the streak appeared at two different locations, and by generating the sky mosaic and tracing back the streaks it was discovered that they came from 1A\,1742-294, an X-ray binary that was in outburst.}
\label{StreakImage}
\end{figure}

These streaks are rare, but have been observed at several locations in the Galactic Center. They were caused by the same source and once the full mosaic was compiled, as shown in Figure \ref{StreakImage}, they were identified as originating from the binary 1A\,1743-294 in outburst during the observations \cite{Hong2016}.

\subsection{Back Reflections}
In BR, the photons strike the backside of the upper mirror of the adjacent shell first then reflect again off the front side of the mirror shell. Because the graze angles of adjacent shells are only slightly different, the photons exit at almost the same angle with which they came from the source. 

Like for the GR, the BR is limited to certain off-axis angles, and at any particular off-axis angle only a few shells contribute at a time. As shown in Figure \ref{GRillustration} the condition for a photon to exit the optics is constrained by the opening angle of the lower shell. For the innermost shell the allowed angles are $\sim 14 \pm 9$\am\ and for the outermost shell it is $\sim 24 \pm 24$\am. The aperture stop further limits these angles, and the geometrical area obtained without including reflection is shown in Figure \ref{GRillustration}.

\begin{figure}[h]
\begin{center}
\includegraphics[width=0.95\textwidth]{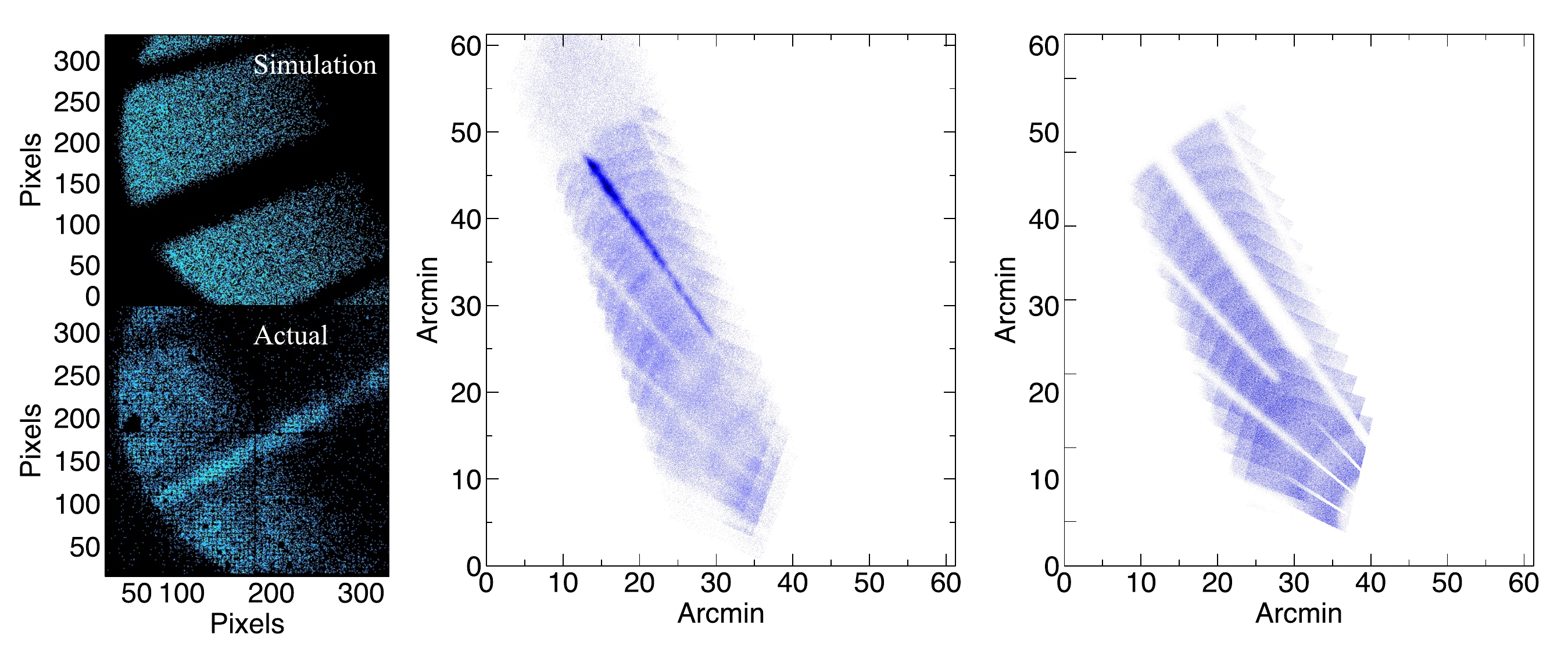}
\end{center}
\caption{\textbf{Left}: Two examples of the GR and BR distribution on the detector from actual NuSTAR data and simulation. The fidelity of the simulation is greater than for the actual, since the detectors were operating at 99\% dead-time with practical exposure times on the order of $\sim 20$\,seconds. \textbf{Middle}: Solar observation, north pole slew. The additional counts apparent at y-axis 50--60\as\ are from  the diffuse cosmic X-ray background. \textbf{Right}: Raytrace of the same observation. The narrow stripes show where the spacers are blocking the X-rays. The broad stripe has no mirror and we see photons in the actual observation at that location, because it happens to be centered such that the flux from the solar X-flare are allowed to pass through. In the bottom right of the image the first 10\am\ are GR.}
\label{sunslew7}
\end{figure}

Reflection from the backside requires very high fluxes to be detected and can not be seen from typical astrophysical sources. This component was, however, observed during the solar observations on 11 October 2014 when a solar X-class flare went off a few hours before the scheduled observations. We show in Figure \ref{sunslew7} an example of how the component looks with the accompanying raytrace simulation (\textit{left}), the full mosaic of the actual observation (\textit{middle}), and its raytrace (\textit{right}). The slew was away from the solar north pole, and the bright streak is the flare entering in through the mirror gap. In the simulations, the mirror gap is larger than in reality; we simulated the as-designed width of the ‘double street’, spacer to spacer (see Figure \ref{optics}), which assumes that the mirror overhang between the spacers is non-focusing due to the surface not being constrained to a conical shape anymore. From imaging data there is some indication that at least part of the mirror overhang is properly focusing, but taken together with the jagged edges of the glass, we have no good way of estimating how much that is. The sharp circular edges are caused by the aperture stop. In the bottom of the mosaic the GR component is just visible as a brighter area before transitioning to the BR. We ran the simulation with a longer exposure time than the actual observation to get better fidelity and more detail in the transition area between GR and BR, and to reduce the errors on the effective area calculation.

\begin{figure}[h]
\begin{center}
\includegraphics[width=0.95\textwidth]{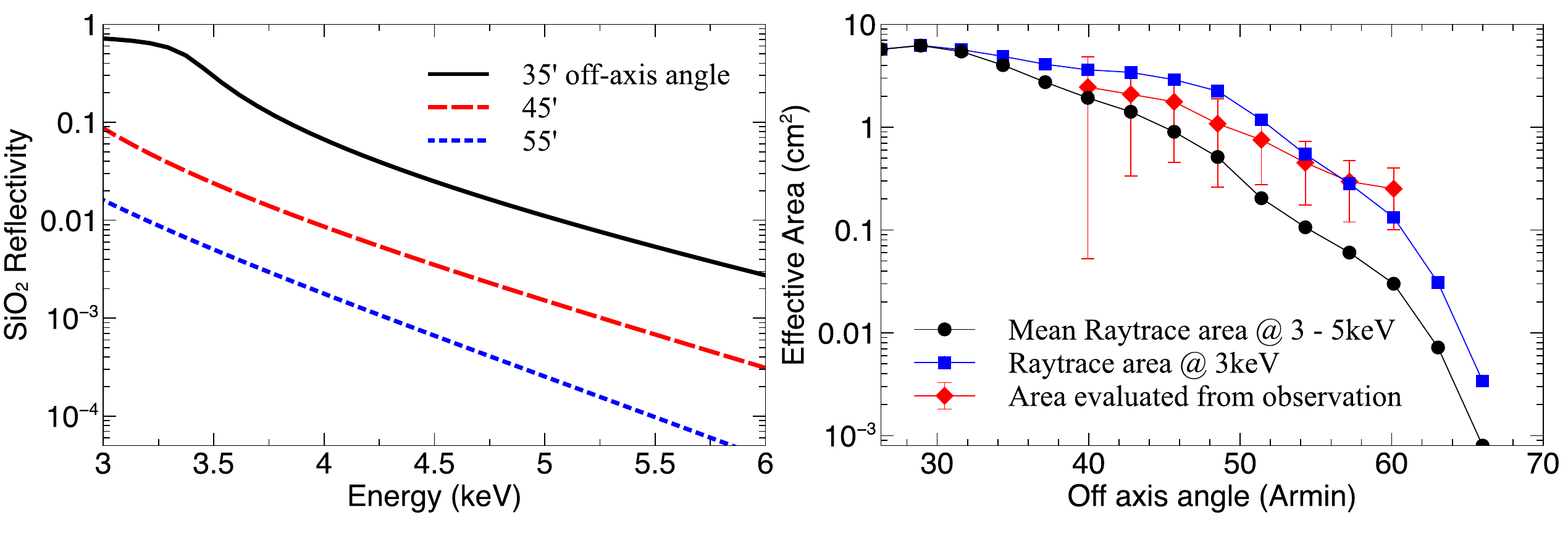}
\end{center}
\caption{\textbf{Left}: The reflectivity curves of a photon reflecting off a 0.21\,mm SiO$_2$ substrate at three different grazing incidence angles. \textbf{Right}: The average effective area between 3--5\,keV for the detector from the raytrace assuming averaged reflection coefficients and effective area obtained from the actual \nustar\ observation of the sun.}
\label{sunslew6}
\end{figure}

To estimate the true effective area of this component we require knowledge of the reflectivity coefficients off the backside of the mirror. We calculated the reflectivity from the inverted multilayer stack with an added 0.21\,mm SiO$_2$ substrate at the top and found that very few photons make it into the stack through the substrate at the angles in question, making the glass substrate the primary contributor. Because of the inefficiency of SiO$_2$ as a reflective surface, only soft photons that can undergo total external reflection have a non-negligible contribution. This causes the BR component to cut off sharply between 3--5\,keV as shown in Figure \ref{sunslew6} (\textit{left}). The mean effective area between 3--5\,keV (black curve in Figure \ref{sunslew6}, \textit{right}) is obtained from the raytrace using the mean SiO$_2$ reflectivities between 3--5\,keV for the first reflection, and the \nustar\ multilayer recipes \cite{Madsen2009} for the second reflection. We did not include the optics thermal cover and detector window Be absorption, and the area has been scaled to the photons falling on the detector area only.

To obtain an independent verification of the effective area we studied the full mosaic of the solar observation. Thanks to the streak, we can extract a flux for the solar flare for the tiles where it was present. The solar flux is extracted from the detector by laying down an area around the streak and using that as the effective area, including all absorption effects in the photon path \cite{Madsen2017}. We then extracted the photon count from the remaining detector, limiting the energy range to be between 3--5\,keV. The effective area is obtained by dividing the detector photon rate with the expected flux, and as shown in Figure \ref{sunslew6} (\textit{right}) there is good agreement between the two estimates. 

The errors on the effective area from the observation mainly comes from evaluating the area of the streak, which is almost certainly narrower than the extraction region laid down on the detector. The jagged edges of the mirror also vary the gap width and cause additional scattering, and we thus estimate a 50\% error on the area. On the raytrace side we assumed the mean reflection values between 3--5\,keV for the backside of the mirrors without weighting it by an input spectrum. This will underestimate the area at higher off-axis angles due to there being more low-energy than high-energy photons in the averaged energy interval. We show in comparison the raytrace effective area at 3\,keV, which as expected shows better agreement with the measured area. 

In comparison, the on-axis effective area in the same energy band is $\sim 500$\,cm$^2$ focused on a small area instead of the entire detector. If we take the typical extraction area to be the PSF Half Power Diameter, which is 1\am\ or 2.9\,mm, then the area of the extraction region is 6.6\,mm$^2$ and the effective area per detector area $\sim 500/6.6 = 75$\,cm$^2$\,mm$^{-2}$. The BR flux covers on average about 70\% of the detector area, which is 1120\,mm$^2$, and so the effective area per detector area for 1\,cm$^2$ of BR is $\sim 1 \times 10^{-3}$\,cm$^2$\,mm$^{-2}$. The source countrate is thus reduced by a factor of $\sim7.5 \times 10^4$ in a comparable typical extraction region and thus completely negligible if caused by typical astrophysical sources.

\section{Stray-Light}
\label{sec:straylight}
\subsection{Primary Stray-Light}
Stray-Light (SL) is the term given to photons that arrive directly at the focal plane without having undergone any reflection or transmission through the optics. 

\begin{figure}[h]
\begin{center}
\includegraphics[width=0.95\textwidth]{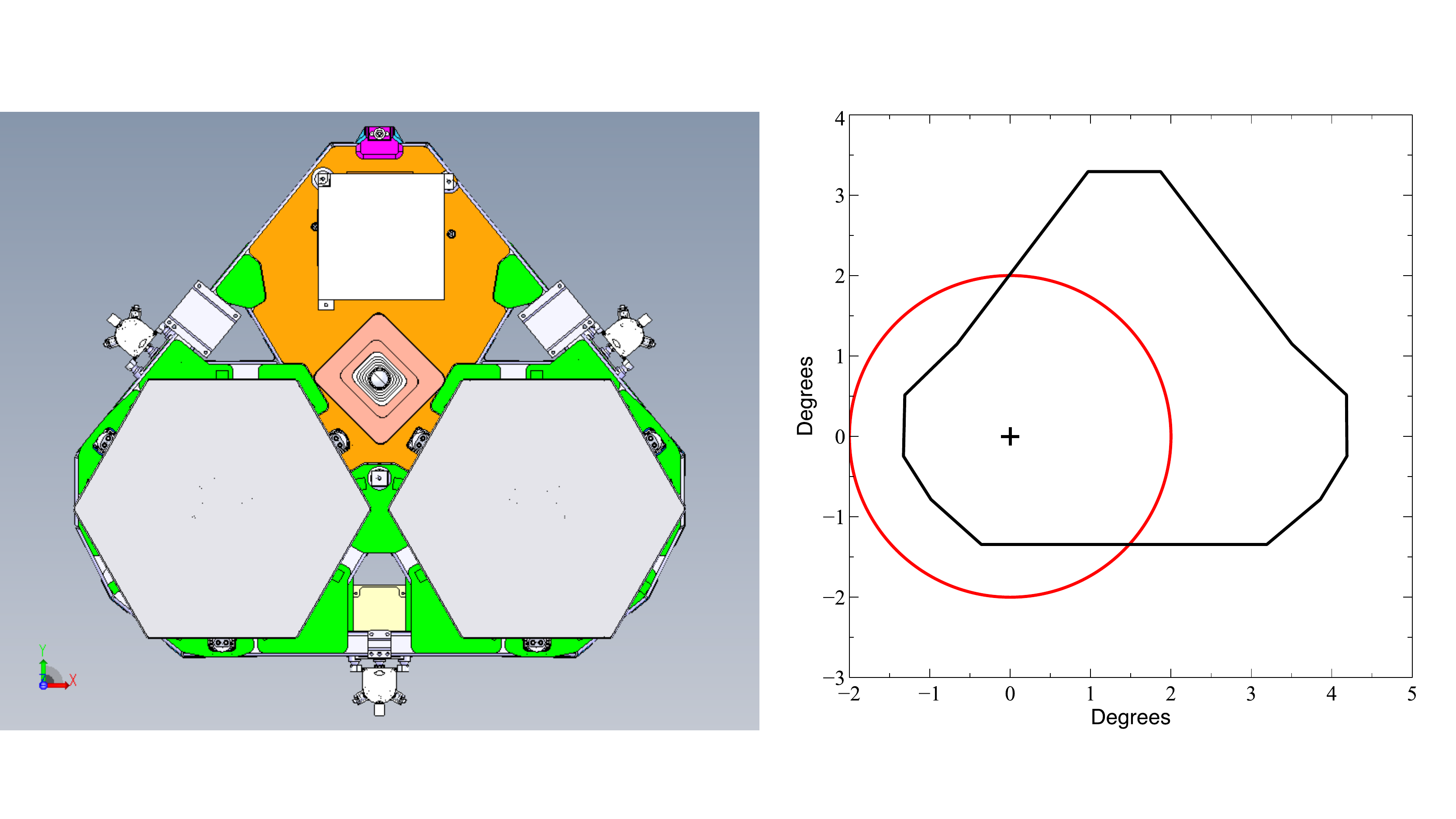}
\end{center}
\caption{\textbf{Left}: The OB as seen from the sky. Hexagonal plates mark the location of the two optics. \textbf{Right}: Projected outline of the OB in degrees on the sky as seen from the detectors. Red circle is the projected opening of the limiting aperture stop for the center of one of the modules. This circle is displaced depending where on the detector one is looking up from, with a max displacement of 2 degrees.}
\label{OBimage}
\end{figure}

The geometry of the SL is determined by the aperture stops, Figure \ref{instrument} (\textit{middle}), and the silhouette of the optical bench. A schematic of the optical bench is shown in Figure \ref{OBimage} and outlines the angular extent of the bench as seen from the detectors. The circle marks the FoV of the sky as seen from the center of one detector up through the aperture stop. Within this FoV there are slices of sky that are not blocked by either the aperture or the bench, and if a source should be located there it may directly reach the focal plane. If a source does happen to occupy that space, it can be blocked by choosing a different Position Angle (PA) of the observatory, which rotates the outline of the bench on the sky. Different areas of the detector see different areas of the sky, and stray-light may occur from sources within 1--4\degree of the observed target. 

\begin{figure}[h]
\begin{center}
\includegraphics[width=0.95\textwidth]{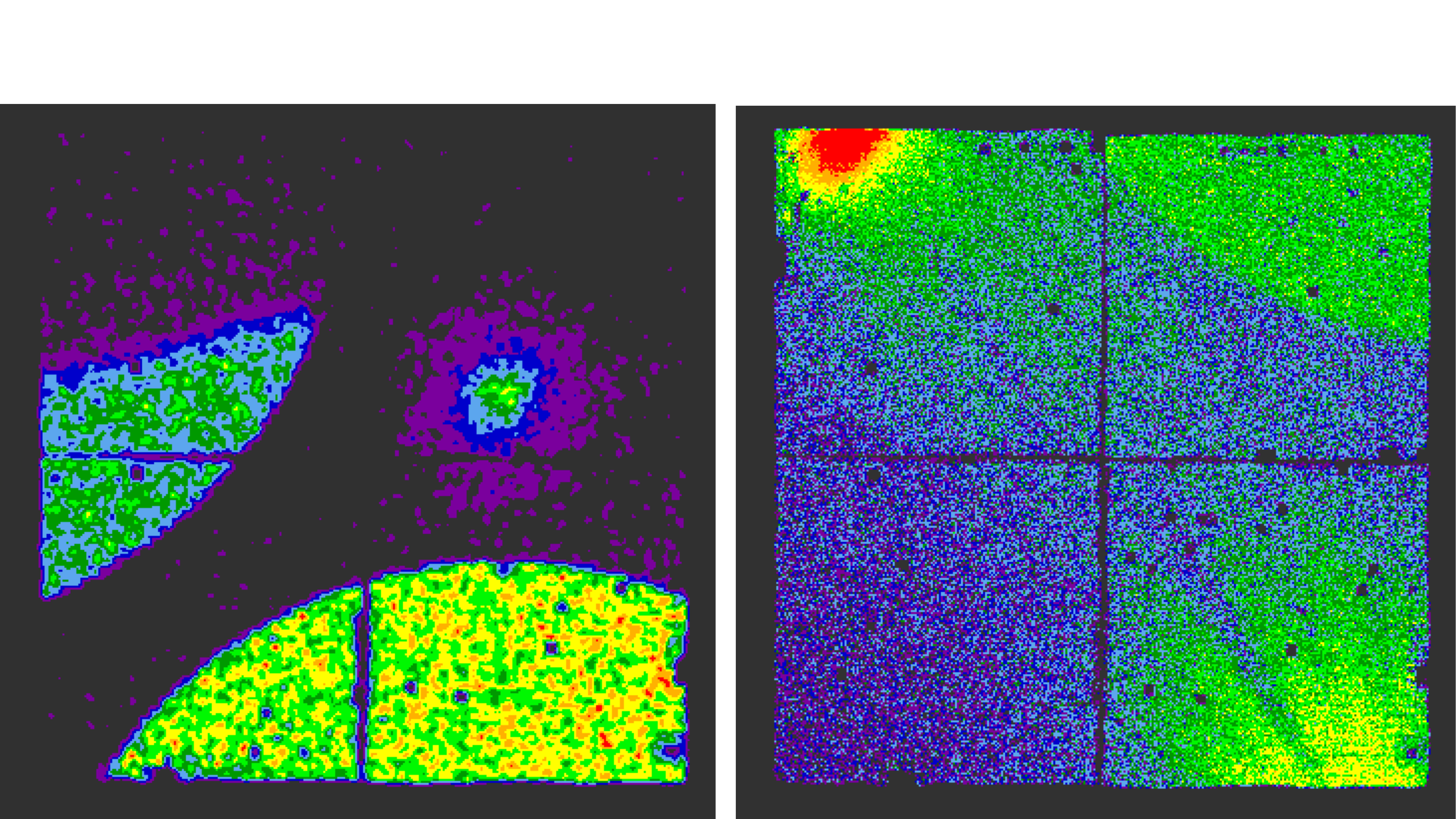}
\end{center}
\caption{\textbf{Left}: SL pattern in \nustar\ observation obsID: 90201027002. The source can be seen in the top right quadrant. The truncation of one crescent is the optical bench blocking the photons. The PA was chosen to allow the source to fall in a SL-free region. The image has been smoothed with a 3 pixel Gaussian. \textbf{Right}: Example of \nustar\ observation obdID: 90201020002 that has both GR and SL.}
\label{SLexample}
\end{figure}

This component places by far the tightest constraints on the planning of observations. The SL can in most cases be completely avoided by picking a suitable PA, but this in turn limits the times a year a target can be observed. Figure \ref{SLexample} shows an example of a careful planning that enable a source to be observed despite multiple SL sources. We also show an example (\textit{left}) where both SL and GR were present to illustrate the visual difference between the two components. The characteristic circular shape of the SL comes from the aperture stop, and because of the simple geometry it is straightforward to predict the location of the SL.

In the rare instances that a science target can not be scheduled to avoid SL, the SL needs to be treated as an additional background. Fortunately, there are a couple of mitigating factors that make analysis straightforward: 
\begin{itemize}
\item Within the illuminated area the spectrum is constant as a function of location. Obtaining a spectrum just requires replacing the mirror effective area with the area covered by the detector. Details on the exact method can be found in \cite{Madsen2017}.
\item Background subtraction is often the greatest concern when dealing with SL. However, due to the unfocused nature of the SL and the typically exponentially declining spectra, the SL spectrum has often fallen far below the internal background at energies where background matters and therefore sometimes be ignored altogether. 
\end{itemize}
With careful analysis most stray-light regions can therefore be dealt with even if they overlap the actual target of the observation.

\subsection{Absorbed Stray-Light}
Similar to the primary SL, absorbed stray-light (ASL) also arrive at the focal plane directly from the source, but the angles are larger, going all the way out to 10\degree, and they do so by transmission through the material of the aperture stops.

The aperture stops were designed to be 1.88\,mm, layered into 0.75 mm Al, 0.13 mm Cu, and 1.0 mm Sn. However, as we will show below, they appear to have been built without the 1.0\,mm of Sn. This allows strong hard spectral sources to transmit through the apertures above $\sim 20$\,keV. Unlike the primary unabsorbed SL, this component is very weak and only a handful of the brightest astrophysical sources (e.g the Crab, Cygnus X-1, and GX\,9+9) have been capable of producing a significant detection. Over five years of operation less than ten observations have been impacted.

\begin{figure}[h]
\begin{center}
\includegraphics[width=0.3\textwidth]{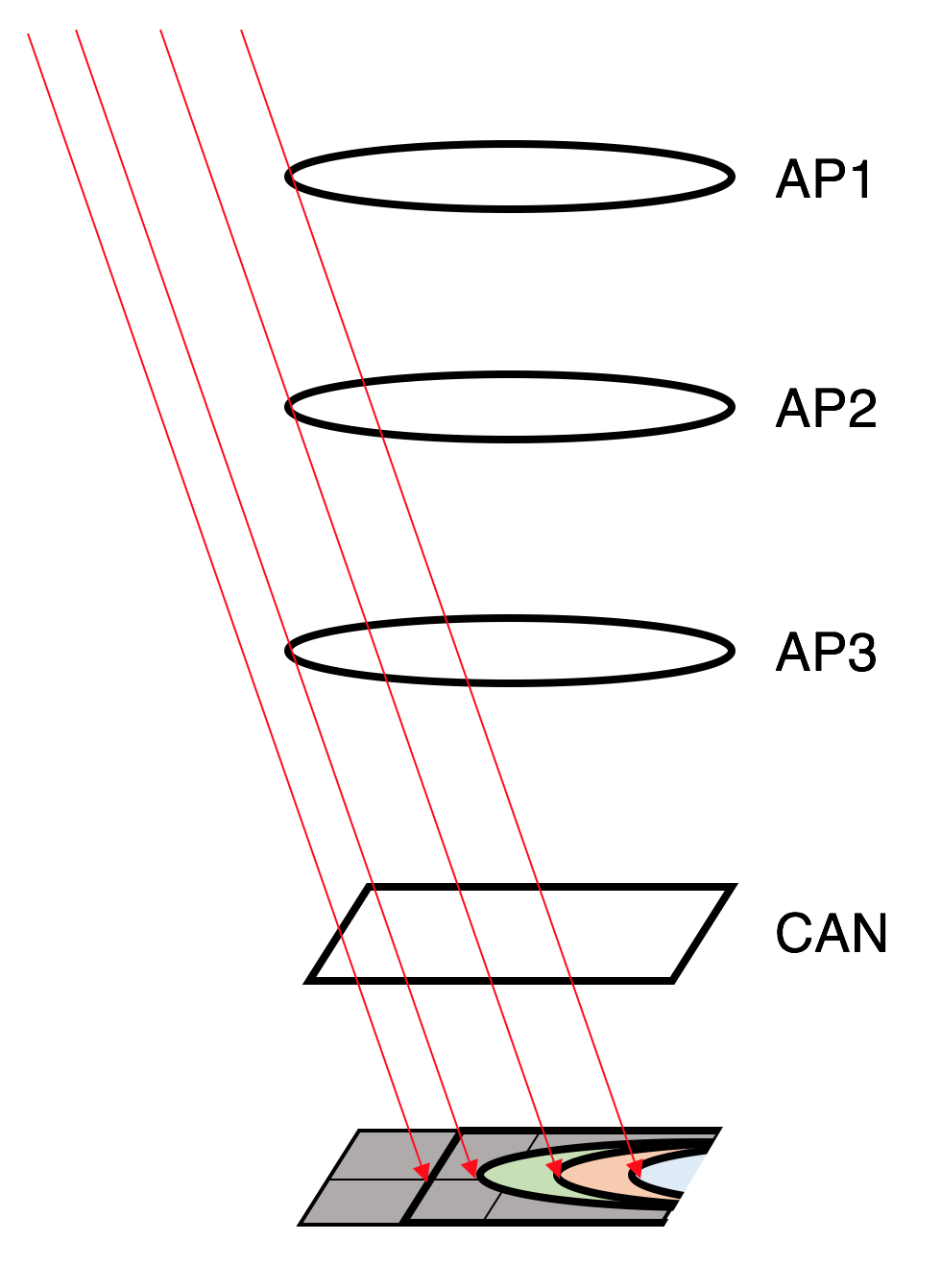}
\end{center}
\caption{Aperture stop schematic. There are three aperture stops, surrounded by 0.13\,mm Cu and 0.75\,mm Al, and a square opening in the detector module acting as the fourth. Rays that enter through the top aperture and hit the detector without transmitting through any of the other apertures is the primary SL (blue circle). Rays that transmit through the first aperture are single absorbed (orange circle), and rays that transmit through the first and second aperture stop double absorbed (green circle).}
\label{ASLschematic}
\end{figure}

Figure \ref{ASLschematic} shows a schematic of the geometry. The top is the limiting aperture, and it leaves a circular SL as illustrated in shade. For strong sources the high energy flux is capable of transmission through the aperture stop and thus photons that have progressed through the first aperture stop and managed to slip through the opening of the second, leave another circle, or crescent, of once absorbed photons. Photons that have transmitted through the first and the second aperture stop material leave a third circle of twice absorbed photons. The top of the detector module is a square opening (`the can') that only allows photons to pass through the opening, no transmission from photons that hit the side. Any photons arriving at angles that are larger than the can's FoV are rejected. Due to their complexity (see Figure \ref{instrument}), we do not model the fixtures that hold the apertures in place, and observations have shown that we can ignore its extent, since the "can" excludes photons that arrive at angles where absorption from the fixture would have been important. Note, however, that a photon may transmit through AP2 or AP3 without having encountered AP1 or AP2. 

\begin{figure}[h]
\begin{center}
\includegraphics[width=\textwidth]{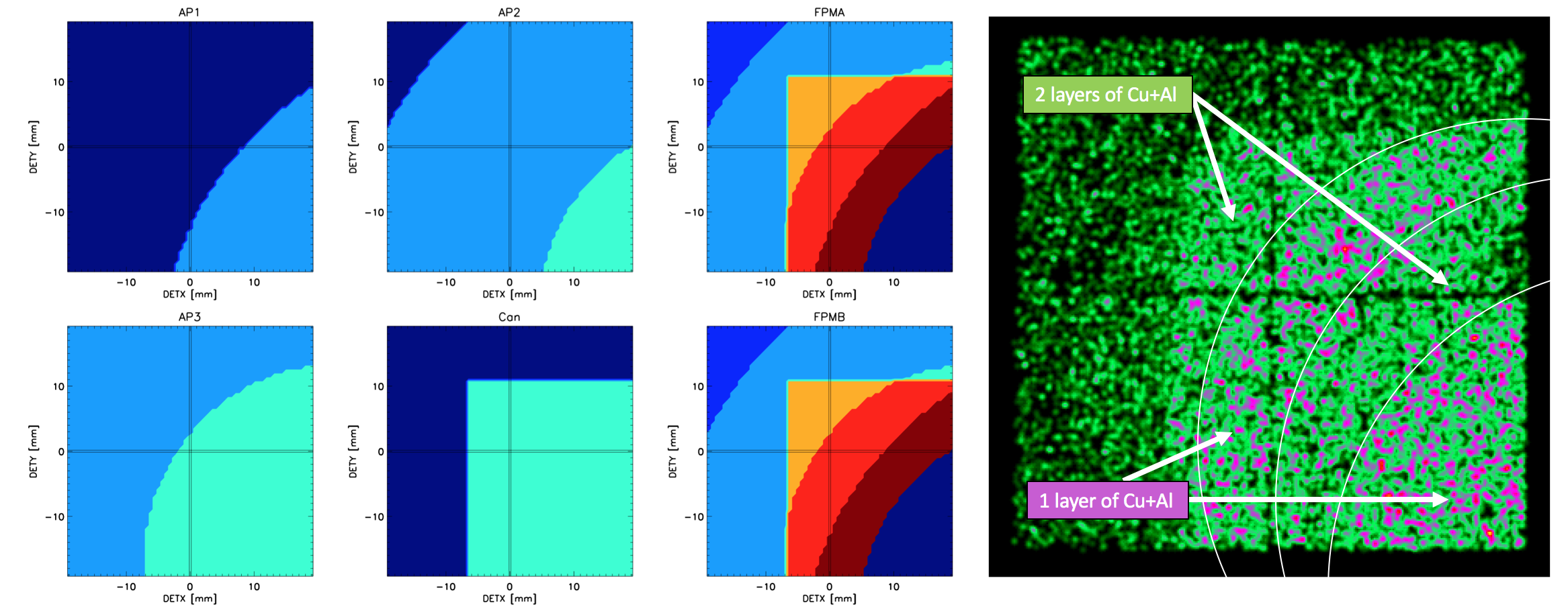}
\end{center}
\caption{\textbf{Left}: Predicted ASL. The four aperture plots individually show the projection of each aperture (AP1-3) and the "can", and the two plots labeled "FPMA" and "FPMB" are the summed aperture images on the two modules (for certain angles the two can be different). In these two latter plots the color shading is a visual aid to distinguish the different regions and should not be taken literally. In the four aperture plots, the smallest (lightest shade, green) crescent shows where SL passes through unabsorbed, and the second largest circle (light blue) where it is absorbed by a single layer of Cu+Al. The darkest region (dark blue) represents the fixture, which we do not model due to its complexity, and observations show that we can ignore its extent, since the "can" excludes photons that arrive at angles where the fixture would have been important, like in this example. \textbf{Right}: Actual ASL. Without the predicted ASL pattern it would be difficult to see where it is once and twice absorbed.}
\label{ASLimage}
\end{figure}

Figure \ref{ASLimage} shows the predicted ASL from Cygnus\,X-1 at an off-axis angle of 3.98\degree\ (\textit{left}), and the actual detector image of the observation (\textit{right}). It is easy to see that there is additional flux on the detectors, but without the predictive plots it is not straightforward to distinguish the boundaries. The flux is therefore not even across the ASL region the way it is for the primary SL, and treating it as an additional background is difficult without precise knowledge of both the actual source spectrum, the layout of the absorbing elements, and their relative occupation underneath the target source.

\begin{figure}[h]
\begin{center}
\includegraphics[width=0.95\textwidth]{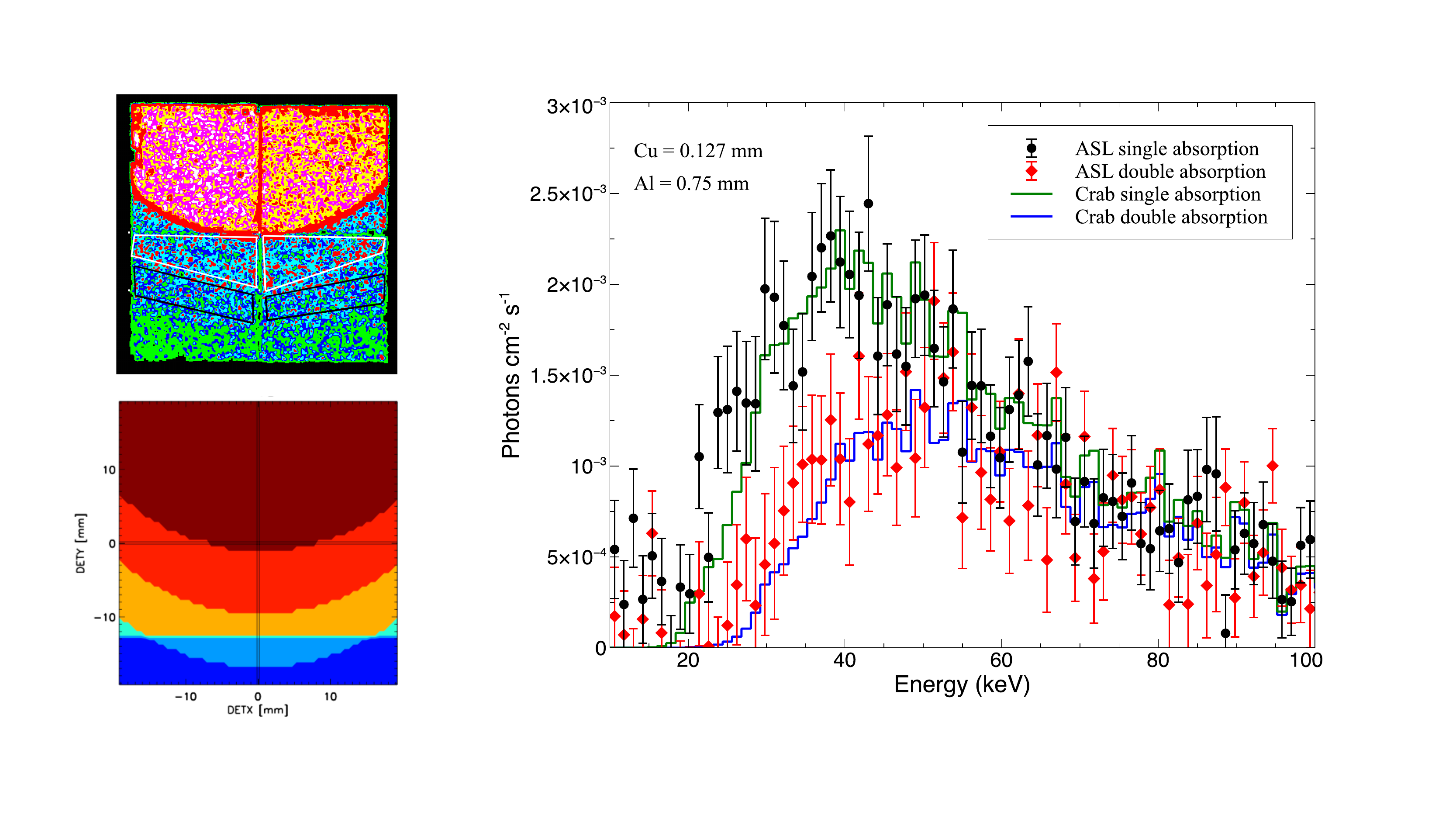}
\end{center}
\caption{\textbf{Left}: Actual and predicted ASL pattern for Crab observation (\nustar\ observation ID: 10110005001) positioned 1.9\degree\ off-axis. The white regions are of single absorbed and black regions double absorbed. \textbf{Right}: The spectra of the single absorbed and double absorbed regions together with the Crab SL spectrum from the top of the detector with the as-built absorption of Cu and Al applied.}
\label{ASLspectrum}
\end{figure}

The ASL spectrum itself can be easily identified because of its characteristic low energy absorption and peak flux at 20--40\,keV. Figure \ref{ASLspectrum} (\textit{right}) shows the Crab spectrum as extracted from the single absorbed and double absorbed regions (\textit{left}). We also extracted the Crab spectrum from the SL region and applied to it the absorption from single and twice absorbed 0.75\,mm Al and 0.13\,mm Cu. This unfortunately shows the absence of the Sn, which would have completely suppressed the spectrum had it been present.

Because of the complicated patterns, albeit predictable, the ASL currently requires specialized, non-standard treatment.

\section{Summary and Discussion}
\label{sec:discussion}
We summarize in Table \ref{summarytable} the different components and the approximate off-axis locations for where they may be observed. All of these artifacts are predictable and most could have been avoided altogether as we will discuss below. We therefore strongly encourage future observatories to investigate each of these components in their optical design. 

The GR are the most difficult to design against. In this paper we used the raytrace approach to investigate the GR and BR because we knew the precise geometrical layout of our instrument and the optical prescription of the multilayer recipes. But in concept designs where a phase-space is being considered there are excellent analytic approaches to help in such an investigation \cite{Spiga2011,Spiga2015}. Some reduction of these components can be achieved by changing the shell spacing and length of the mirrors, but as mentioned it can not completely eliminate the GR. To further reduce the GR, baffling of some sort is required. For \nustar\ we designed a baffle made of Invar to be placed in front of the optics that extended the height of the mirror shells to reject photons coming in at a range of off-axis angles. However, due to launch mass constraints it was not implemented. A similar design was used for \xmm\ and with it they are able to reduce about 80\% of the GR flux \cite{Chambure1999}. When the shell spacing is larger, as it is for \chandra\, the baffling can be done within the optics \cite{Gaetz2000,Cusumano2007}.

To reduce the SL and background, \nustar\ designed a deployable optical shield, which would have increased the angular extent of the optical bench, and although built it was never mounted due to pre-launch scheduling constraints. An alternative approach for soft X-ray instruments would be to shroud the optical path, but in the case of \nustar\ that was not a feasible approach since the amount of shielding required to stop the high energy flux would have been prohibitive. For future hard X-ray instrumentation with long focal lengths, careful thought must therefore be put into the design of baffling and apertures down the length of the optical path.

For the ASL, the inclusion of Sn in the apertures as designed would have been sufficient to eliminate the component. Elimination of the `streak' could have been achieved simply by blocking the gap between mirrors segments.

Overall, the artifacts have resulted in some scheduling constraints for the planning of targets. Because of the GR there is a region around bright sources ($< $1\am) where a target of an observation must be of a certain brightness not to be affected by GR. But even then, as demonstrated, the GR often leaves contamination free regions. When allowing for different observing angles, and waiting out the contamination of bright transients, the majority of targets can be observed without significant issues.

\acknowledgments % equivalent to \section*{ACKNOWLEDGMENTS} 
This work was supported under NASA Contract No. NNG08FD60C, and made use of data from the NuSTAR mission, a project led by the California Institute of Technology, managed by the Jet Propulsion Laboratory, and funded by the National Aeronautics and Space Administration. We thank the NuSTAR Operations, Software and Calibration teams for support with the execution and analysis of these observations. This research has made use of the NuSTAR Data Analysis Software (NuSTARDAS) jointly developed by the ASI Science Data Center (ASDC, Italy) and the California Institute of Technology (USA).   
 
% References
\bibliography{bib} % bibliography data in report.bib
\bibliographystyle{spiebib} % makes bibtex use spiebib.bst

\section{Biographies}
\subsection*{Kristin K. Madsen}
Kristin K. Madsen received her Ph.D in astrophysics from the University of Copenhagen in 2007. She has over 10 years of experience in X-ray optics and astrophysical research and is currently a staff scientist at the California Institute of Technology as one of the principle instrument and mission scientists for \nustar. Her research interests include active galactic nuclei, X-ray binaries, pulsar wind nebulae, and hard x-ray optical design.
\subsection*{William W. Craig}
William Craig received his Ph.D. in astrophysics from U.C. Berkeley in 1994. He has over 20 years of experience in astrophysical instrumentation and research, working with both optical and detector systems and has worked on numerous space missions including \textit{XMM-Newton, CHIPS, GLAST, NuSTAR and ICON}, as well as a number of balloon-borne instruments.  He is currently the project manager for the \textit{Ionospheric Connections Explorer (ICON)} at UC Berkeley.
\subsection*{Finn E. Christensen} Finn christensen received his Ph.D in 1982 from the Danish technical university. His current position is senior staff scientist at DTU-space and he has more than 30 year experience in developing, studying and calibrating X-ray optics for space research. His current research revolves around developing and calibrating the coatings for the ATHENA mission.
\subsection*{Karl W. Forster}
Karl Forster is a graduate of the University of Leicester and the University of Hertfordshire in the UK, and was awarded a PhD in astronomy from Columbia University. His research interests include high energy emission from active galactic nuclei and has developed and managed science operations for NASA missions \textit{GALEX} and \nustar.
\subsection*{Brian W. Grefenstette}
Brian Grefenstette is a research scientist at the California Institute of Technology. He is one of the principle mission scientists for \nustar, working on the calibration of the focal plane detectors. His science interests include nuclear astrophysics, X-ray binaries, supernovae, and supernova remnants.
\subsection*{Fiona A. Harrison}
Fiona A. Harrison received her Ph.D. in physics from the University of California, Berkeley in 1993. She is the California Institute of Technology (Caltech) Benjamin M. Rosen Professor of Physics, and the Kent and Joyce Kresa Leadership Chair of the Division of Physics, Mathematics and Astronomy. Her research is focused on the study of energetic phenomena ranging from gamma-ray bursts, black holes on all mass scales, to neutron stars and supernovae. She is the principal investigator for \nustar. 
\subsection*{Hiromasa Miyasaka}
Hiromasa Miyasaka received his Ph.D in Physics from Saitama University (Japan) in 2000. He has worked on numerous space missions including French-Brazilian micro-satellite, \textit{Suzaku, STEREO, NuSTAR} and Parker Solar Probe. He is involved in development on various radiation detectors as well as custom low-noise application specific integrated circuits (ASIC). He is currently a staff member of \nustar\ Science Operation Center at Caltech.
\subsection*{Vikram Rana} 
Vikram Rana is a staff scientist in Space Radiation Laboratory at Caltech. He obtained his Ph.D. in Astronomy and Astrophysics from Tata Institute of Fundamental Research (TIFR), India, in 2006. His research experience include development of X-ray optics for Astrosat and focal plane radiation detectors for NuSTAR missions. He is also interested in studying X-ray emissions from Ultraluminous X-ray sources, X-ray Binaries and Cataclysmic Variables.

\begin{table}
\centering
\caption{Summary of Artifacts}
\begin{tabular}{|l|c|c|}
\hline
Component & minimum angle & maximum angle \\
\hline
GR upper reflection (\am) & 2 & 10 \\
GR lower reflection (\am) & 5 & 30 \\
BR reflection (\am) & 15 & 65 \\
Streak (\am) & 37 & 65 \\
SL (\degree) & 1 & 4 \\
ASL (\degree) & 1 & 10 \\
\hline
\end{tabular}
\label{summarytable}
\end{table}

\end{document}